\documentclass[nofootinbib,a4paper,fleqn,aps,prd,10pt,superscriptaddress,reprint,showkeys,showpacs]{revtex4-1}

\usepackage{times}
\usepackage{amsfonts}
\usepackage{amsmath}
\usepackage{amssymb}
\usepackage{natbib}
\usepackage{graphicx}
\usepackage{enumitem}
\usepackage{xcolor}
\usepackage[colorlinks=true,linkcolor=blue]{hyperref}
\usepackage[outdir=./]{epstopdf}
\usepackage[normalem]{ulem}

\def\aap{A\&A}
\def\aj{AJ}
\def\apj{ApJ}
\def\apjl{ApJL}
\def\apjs{ApJS}
\def\apss{Astrophysics and Space Science}
\def\jcap{JCAP}
\def\mnras{MNRAS}
\def\na{New Astronomy}

\def\physrep{Physics Reports}

\begin{document}

\title{{\bf Turnaround radius in $\Lambda$CDM}, and dark matter cosmologies with shear and vorticity.
}

\author{Antonino~\surname{Del Popolo}}%
\affiliation{%
Dipartimento di Fisica e Astronomia, University of Catania, Viale Andrea Doria 6, 95125, Catania, Italy
}
\affiliation{%
Institute of Astronomy, Russian Academy of Sciences, 119017, Pyatnitskaya str., 48 , Moscow 
}
\affiliation{%
INFN sezione di Catania, Via S. Sofia 64, I-95123 Catania, Italy
}
\email[Corresponding author: ]{adelpopolo@oact.inaf.it}

\author{Man Ho~\surname{Chan}}%
\affiliation{Department of Science and Environmental Studies, The Education University of Hong Kong, Tai Po, New Territories, Hong Kong }
\email[]{chanmh@eduhk.hk}

\author{David F.~\surname{Mota}}%
\affiliation{Institute of Theoretical Astrophysics, University of Oslo, P.O Box 1029 Blindern, N-0315 Oslo, Norway }
\email[]{chanmh@eduhk.hk}



\label{firstpage}

\date{\today}

\begin{abstract}
We determine the relationship between the turnaround radius, $R_{\rm t}$, and mass, $M_{\rm t}$, in $\Lambda$CDM, and in dark energy scenarios, using an extended spherical collapse model taking into account the effects of shear and vorticity.
We find a more general formula than that usually described in literature, showing a dependence of $R_{\rm t}$ from shear, and vorticity.
The $R_{\rm t}-M_{\rm t}$ relation differs from that obtained not taking into account shear, and rotation, 
especially at galactic scales, differing $\simeq 30\%$ from the result given in literature. 
This has effects on the constraint of the $w$ parameter of the equation of state. 
We compare the $R_{\rm t}-M_{\rm t}$ relationship obtained for the $\Lambda$CDM, and  different dark energy models to that obtained in the $f(R)$ modified gravity (MG) scenario. The $R_{\rm t}-M_{\rm t}$ relationship in $\Lambda$CDM, and dark energy scenarios are tantamount to the prediction of the $f(R)$ theories. {Then, the $R_{\rm t}-M_{\rm t}$ relationship is not a good probe to test gravity theories beyond Einstein's general relativity.}
\end{abstract}



\maketitle

\section{Introduction}

In the past decades the $\Lambda$CDM model has successfully passed many tests \citep{Komatsu2011,Planck2014,
DelPopolo2013}, deserving to be dubbed "the standard model of cosmology". One of the assumption on which modern cosmology is based, is that General Relativity is the correct theory of gravitation. In this case, observations indicate the existence of a larger content of mass-energy than predicted \cite{DelPopolo2007,DelPopolo2014aa,Bull2016}.
The mass-energy of the universe is dominated by non-baryonic and non-relativistic particles, indicated as "cold dark matter" \cite{DelPopolo2014}, and a second component, dubbed "dark energy" (DE), a fluid with exotic properties, like that of having negative pressure, and giving rise to the accelerated expansion of the universe.
In its simplest form, and in the $\Lambda$CDM model, DE is represented by the cosmological constant 
$\Lambda$.  

Nevertheless the success of the $\Lambda$CDM model \cite{Spergel2003,Komatsu2011,DelPopolo2007}, precision data are revealing drawbacks, and tensions both at large scales \citep{Eriksen2004,Schwarz2004,Cruz2005,Copi2006,Macaulay2013,Planck2014_XVI,Raveri2016}, and at small ones 
\cite{Moore1999,deBlok2010,Ostriker2003,BoylanKolchin2011,DelPopolo2014a,DelPopolo2014d,DelPopolo2017a}. 

To start with, the particles that should constitute the DM has never been observed \cite{Klasen2015}, despite a large number of indirect evidences from small to large scales  \cite{Einasto2001,Bertone2005,Bouchet2004,Kilbinger2015}, and a large campaign of direct and indirect searches \cite{Bertone2005,Klasen2015, DelPopolo2014}. 
{In particular, in a recent direct-detection experiment, XENON1T \citep{Aprile2018}, no significant excess over the background was found. For indirect searches, a large parameter space for annihilating dark matter was ruled out based on the radio data of the Andromeda galaxy \citep{Chan2019} and the gamma-ray data of the A2877 and Fornax clusters \citep{Chan2017}.}

The so called "small scale problems" of the $\Lambda$CDM \cite{DelPopolo2017a} are plaguing the model, and  several recipes, cosmological \citep{Zentner2003}, different nature of the dark matter particles \citep{Colin2000,Goodman2000,Hu2000,Kaplinghat2000,Peebles2000,SommerLarsen2001}, MG theories, e.g., $f(R)$ \citep{Buchdahl1970,Starobinsky1980}, $f(T)$ \citep{Bengochea2009,Linder2010,Dent2011,Zheng2011}, MOND \citep{Milgrom1983}
and astrophysical \cite{Brooks2013,Onorbe2015,DelPopolo2017a}, have been proposed to solve those problems.

Apart the issues related to the DM component of the $\Lambda$CDM model, the cosmological constant $\Lambda$ 
suffers from the ``cosmological constant fine tuning problem", and the ``cosmic  coincidence problem" \cite{Astashenok2012,Velten2014,Weinberg1989}. 
 
The quoted issues motivated the investigations of other explanations, and models to clarify the universe accelerated expansion. These alternative models generate the DE effects through additional matter fields (e.g., quintessence \citep{Copeland2006}), or MG models\citep{Horndeski1974,Milgrom1983,Zwiebach1985,Moffat2006,Nojiri2005,Bekenstein2010,DeFelice2010,Linder2010,
Milgrom2014,Lovelock1971,Horava2009,Rodriguez2017,Horndeski1974,Deffayet2010,Deffayet2010}. In some cases, the quoted theories tried to explain the accelerated expansion as the manifestation of extra dimensions, or higher-order corrections effects, as in the Dvali-Gabadadze-Porrati (DGP) model \cite{Dvali2000} and in $f(R)$ gravity. 
Disentangle between the plethora of models is not an easy task. The solution of the problem, or at least a better understanding of the same, may come from future surveys like:
Euclid\footnote{\url{http://www.euclid-ec.org}}, JDEM\footnote{\url{http://jdem.lbl.gov/}},
SKA\footnote{\url{https://www.skatelescope.org}}, LSST\footnote{\url{https://www.lsst.org}}, or
from new studies of the CMB \citep{Battye2018a,Battye2018c}.

DE and MG effects can be probed with structure formation. By means of hydro-dynamical simulations, one can obtain observables (e.g., splashback radius \cite{Adhikari2018}, the halo profile, the TAR \cite{Bhattacharya2017,Lopes2018}, or the mass-temperature relation (MTR) \cite{Hammami2017}) which can be directly compared with observations. In a recent paper \citep{DelPopolo2019}, we showed that the MTR relation cannot be used to put constraints on MG theories, since the $\Lambda$CDM model gives similar predictions to that of the MG theories.

Recently, the turnaround radius (TAR)\footnote{The TAR is defined as the distance from the center of a structure to the surface at which radial velocity is zero. The majority of works studies the maximum TAR, defined as the radius at which the radial acceleration is null, $\ddot{r} =0$.} has been proposed as a promising way to test cosmological models \citep{Lopes2018}, DE, and disentangle between $\Lambda$CDM model, DE, and MG models \citep{Pavlidou2014,Pavlidou2014a,Faraoni2015,Bhattacharya2017,Lopes2018,Lopes2019}. The attention on TAR increased when was shown that MG can affect the maximum TAR (see \citep{Lopes2019}).
According to some authors (e.g., \citep{Pavlidou2014}), the TAR is a well-defined, and unambiguous boundary of a structure, in the spherical collapse model (SCM), simulations, analytic calculations, and is clean from baryons physics. Concerning the last point, contrarily to \citep{Pavlidou2014}, we already showed in \citep{DelPopolo2013a,DelPopolo2013b,Pace2014,Mehrabi2017,Pace2019} that the non-linear equation driving the evolution of the overdensity contrast depends on shear, and vorticity, which conversely depends from the mass of the forming structure, and the way it forms. In the following, we will also show that the TAR also depends from dynamical friction. All the papers dealing with the determination of TAR, usually find it from some metric, and equations not taking into account the real physics of structure formation. Then, we will show in this paper, that TAR depends from baryons physics. 

\citep{Pavlidou2014} calculated the TAR for $\Lambda$CDM, and \citep{Pavlidou2014a} did the same for smooth DE. 
{\citep{Lee2016} proposed to use the zero velocity surface of large structures to look for 
the violation of the maximum upper bound of $R_{\rm t}$. 
%
}

In MG theories \citep{Capozziello2019} found a general relation for the maximum TAR in $f(R)$ theories, and \citep{Faraoni2015} found a method to get the same quantities in generic gravitational theories. 

In the present paper, we use an extended spherical collapse model (ESCM) introduced, and adopted in   \citep{DelPopolo2013,DelPopolo2013a,Pace2014,Mehrabi2017,Pace2019}. The ESCM takes into account the effect of a non null shear and vorticity on the collapse, to show how the TAR is changed. 
As we have already shown, shear and vorticity change the typical parameters of the SCM \citep{DelPopolo2013}, the mass function \citep{DelPopolo2013,DelPopolo2013a,Pace2014,Mehrabi2017}, the two-point correlation function \citep{DelPopolo1999}, and the weak lensing peaks \citep{Pace2019}. We will find a more general formula than that obtained by \citep{Pavlidou2014} for smooth DE, showing that shear, and vorticity, give rise to a smaller TAR especially at the galactic mass scales. Then, we will find the TAR - mass relation for the $\Lambda$CDM model, and several dynamical DE models, and compare our results to the prediction of the $R_{\rm t}$-$M_{\rm t}$ relation of \citep{Lopes2018} for $f(R)$ models. The result of the comparison shows that the $R_{\rm t}$-$M_{\rm t}$ relation of the $\Lambda$CDM model has a very similar behavior to that of the $f(R)$ models.
This makes impossible to disentangle between the MG results and those of GR.
A similar situation happens for the DE models.

The paper is organized as follows. 
Section~\ref{sect:model} describes the model used to derive the $R_{\rm t}$-$M_{\rm t}$ relation in $\Lambda$CDM, and smooth DE cosmologies.  
Section~\ref{sect:Results} is devoted to the presentation and the discussion of our results.  
Section~\ref{sect:conclusions} is devoted to conclusions.

%
%
%


\section{The Model}\label{sect:model}

The SCM introduced by \cite{Gunn1972}, extended and improved in several following papers \citep{Fillmore1984,Bertschinger1985,Hoffman1985,Ryden1987,Subramanian2000,
Ascasibar2004,Williams2004}, is a very popular method to study analytically the non-linear evolution of perturbations of DM and DE. 
Shortly, the model describes the evolution of a spherical symmetric overdensity, how it decouples from the Hubble flow, reach a maximum radius, dubbed TAR, and finally collapse and virialize\footnote{
In the virialization process, the collapse kinetic energy is converted into random motions.}.
\cite{Gunn1972} is a very simple SCM model just considering the radial collapse of the structure. It does not take account of tidal angular momentum \citep{Peebles1969,White1984}, random angular momentum \citep{Ryden1987,Ryden1988,Williams2004}, dynamical friction (\citep{AntonuccioDelogu1994,Delpopolo2009}), etc. The way to take account of angular momentum was studied in several papers\citep{Ryden1987,Gurevich1988a,Gurevich1988b,White1992,Sikivie1997,Nusser2001,Hiotelis2002,  
LeDelliou2003,Ascasibar2004,Williams2004,Zukin2010}\footnote{Particles angular momentum is distributed in random directions in order the mean angular momentum at any point in space is zero \cite{White1992,Nusser2001}}, that of dynamical friction was studied in \citep{AntonuccioDelogu1994,Delpopolo2009}, while \citep{Hoffman1986,Hoffman1989,Zaroubi1993} discussed the role of shear in the gravitational collapse.

The SCM 
with negligible DE perturbations was studied by 
\citep[see, e.g.][]{Bernardeau1994,Bardeen1986,Ohta2003,Ohta2004,Basilakos2009,Pace2010,Basilakos2010} 
while in \citep[see][]{Mota2004,Nunes2006,Abramo2007,Abramo2008,Abramo2009a,Abramo2009b,Creminelli2010,Basse2011,Batista2013}
the effects of the DE fluid perturbation were taken into account.  
\citep{Pettorino2008,Wintergerst2010a,Tarrant2012} extended the SCM model to coupled DE models, and 
\citep{Pettorino2008,Pace2014} to extended DE (scalar-tensor) models.

\cite{DelPopolo2013a,DelPopolo2013b} studied the effects of shear and rotation in smooth DE models.
The effects of shear and rotation were investigate in \cite{DelPopolo2013a,DelPopolo2013b} for smooth DE models, \cite{Pace2014b} in clustering DE cosmologies, and 
\cite{DelPopolo2013c} in Chaplygin cosmologies.

\subsection{The ESCM}
Here, we show how the evolution equations of $\delta$ in the non-linear regime can be obtained. 
Those equations were obtained by \cite{Bernardeau1994,Padmanabhan1996,Ohta2003,Ohta2004,Abramo2007,Pace2010} in the spherical and ellipsoidal collapse models scheme. We assume the equation of state $P=w\rho c^2$ for the fluid, and the Neo-Newtonian expressions for the relativistic Poisson equation, the Euler, and continuity equations \citep{Lima1997}
\begin{eqnarray}
  \frac{\partial\rho}{\partial t}+\nabla_{\vec{r}}\cdot(\rho\vec{v})+
  \frac{P}{c^2}\nabla_{\vec{r}}\cdot\vec{v} & = & 0 \label{eqn:cnpert}\;,\\
  \frac{\partial\vec{v}}{\partial
    t}+(\vec{v}\cdot\nabla_{\vec{r}})\vec{v}+
  \nabla_{\vec{r}}\Phi +{\bf \frac{c^2}{c^2 \rho+P} \nabla P} & = & 0\;, \label{eqn:enpert}\\
  \nabla^2\Phi-4\pi G\left(\rho+\frac{3P}{c^2}\right) & = & 0\;,\label{eqn:pnpert}
\end{eqnarray}
where we indicated with $\vec{r}$ the physical coordinate, with $\Phi$ the Newtonian gravitational potential, and the velocity in three-space is given by $\vec v$.

%
%

The perturbations equations, using comoving coordinates $\vec{x}=\vec{r}/a$ are given by
\begin{eqnarray}
 \dot{\delta}+(1+w)(1+\delta)\nabla_{\vec{x}}\cdot\vec{u} & = & 0\;, \label{eq:pertCont}\\
 \frac{\partial \vec{u}}{\partial t}+2H\vec{u}+(\vec{u}\cdot\nabla_{\vec{x}})\vec{u}+\frac{1}{a^2}\nabla_{\vec{x}}\phi &
 = & 0\;,\label{eq:pertEuler} \\
 \nabla_{\vec{x}}^2\phi-4\pi G(1+3w)a^2\bar{\rho}\delta & = & 0\;, \label{eq:pertPois}
\end{eqnarray}
being $\vec{u}(\vec{x},t)$ the comoving peculiar velocity, and $H(a)$ is the Hubble function.
{\bf Note that in Eq. (\ref{eq:pertEuler}) the pressure terms are zero, because 
in the top-hat, with $\rho=\overline{\rho} +\delta \rho$, and $P=\overline{P} +\delta P$ the gradient is zero.}

The non-linear evolution equation is obtained combining the previous equations, which in the case $w=0$ (dust)
%
is given by:
\begin{equation}\label{eqn:nleq11}
\begin{split}
\ddot{\delta}+2H\dot{\delta}-\frac{4}{3}\frac{\dot{\delta}^2}{1+\delta}-
4\pi G\bar{\rho}\delta(1+\delta)-\\
(1+\delta)(\sigma^2-\omega^2) & = 0\;
\end{split}
\end{equation}
Eq. (\ref{eqn:nleq11}) is Eq.~41 of \cite{Ohta2003}, and a generalization of Eq.~7 of \cite{Abramo2007} to the case of a non-spherical configuration of a rotating fluid.

In Eq. (\ref{eqn:nleq11}), $\sigma^2=\sigma_{ij}\sigma^{ij}$, and $\omega^2=\omega_{ij}\omega^{ij}$ are the shear, and rotation term, respectively. The shear term is related to a symmetric traceless tensor, dubbed shear tensor, while rotation term is related to an antisymmetric tensor, given by  
\begin{eqnarray}
 \sigma_{ij} & = & \frac{1}{2}\left(\frac{\partial u^j}{\partial x^i}+\frac{\partial u^i}{\partial x^j}\right)
 -\frac{1}{3}\theta\delta_{ij}\;, \\
 \omega_{ij} & = & \frac{1}{2}\left(\frac{\partial u^j}{\partial x^i}-\frac{\partial u^i}{\partial x^j}\right)\;.
\label{eq:tens}
\end{eqnarray}
being $\theta=\nabla_{\vec{x}}\cdot\vec{u}$ the expansion.

In terms of the scale factor, $a$, 
Eq. (\ref{eqn:nleq11}) takes the form
\begin{equation}\label{eqn:wnldeq}
 \begin{split}
  \delta^{\prime\prime}+\left(\frac{3}{a}+\frac{E^\prime}{E} \right)
\delta^\prime-\frac{4}{3}\frac{\delta^{\prime 2}}{1+\delta}-
  \frac{3}{2}\frac{\Omega_{\mathrm{m},0}}{a^5 E^2(a)}\delta(1+\delta)-&\\
  \frac{1}{a^2H^2(a)}(1+\delta)(\sigma^2-\omega^2)&=0\;,
 \end{split}
\end{equation}
where $\Omega_{m,0}$ is DM density parameter at $t=0$ ($a=1$), and $E(a)$ is given by
\begin{equation}\label{eq:e}
E(a)=\sqrt{\frac{\Omega_{\mathrm{m},0}}{a^3}+\frac{\Omega_{\mathrm{K},0}}{a^2}+ \Omega_{\mathrm{Q},0}g(a)}\;,
\end{equation}
being 
\begin{equation}
g(a)=\exp{\left(-3\int_1^a\frac{1+w(a')}{a'}~da^{\prime}\right)}\;.
\end{equation}

{
Recalling that $\delta=\frac{2GM_{\rm m}}{\Omega_{m,0} H^2_0}(a/R)^3-1$, where $R$ is the effective perturbation radius and inserting into 
Eq.~(\ref{eqn:nleq11}), we get

\begin{eqnarray} \label{eqn:wnldeq0}
 \ddot{R} &=& -\frac{GM_{\rm m}}{R^2} - \frac{GM_{\rm de}}{R^2}(1+3w_{\rm de})-\frac{\sigma^2-\omega^2}{3}R= \nonumber\\
& & 
-\frac{GM_{\rm m}}{R^2} - \frac{4 \pi G \bar{\rho_{\rm de}} R}{3}(1+3w_{\rm de})-\frac{\sigma^2-\omega^2}{3}R\,,
\end{eqnarray}
{being $M_{\rm m}= \frac{4 \pi R^3}{3} (\bar{\rho}+\delta \rho)$ and $M_{\rm de}$ is the mass of the dark-energy component enclosed in the volume, $\bar{\rho}_{\rm de}$, and $w_{\rm de}$ being respectively its background density and equation-of-state \citep{Fosalba1998a,Engineer2000,Ohta2003,Pace2019}.
$M_{\rm m}$, as shown, contains background and perturbation.}

%
%
Eq. (\ref{eqn:wnldeq0}), for $w=-1$, becomes 
\begin{equation} \label{eqn:spher0}
 \ddot{R}=-\frac{GM_{\rm m}}{R^2} -\frac{\sigma^2-\omega^2}{3}R+ \frac{\Lambda}{3} R\
\end{equation}
and is similar to the usual expression for the SCM with angular momentum \citep[e.g.][]{Peebles1993,Nusser2001,Zukin2010}, and cosmological constant:
\begin{equation}
\frac{d^2 R}{d t^2}= -\frac{GM}{R^2} +\frac{L^2}{M^2 R^3}+\frac{\Lambda}{3}R= -\frac{GM}{R^2} + \frac{4}{25} \Omega^2 R+ \frac{\Lambda}{3}R,
\label{eqn:spher}
\end{equation}
The term $\frac{4}{25} \Omega^2 R$ comes from the expression of angular momentum, $L=I \Omega$, and the 
momentum of inertia of a sphere, $I=2/5 M R^2$.

{
In the simple case of a uniform rotation with angular velocity $\Omega=\Omega_{\rm z} {\bf e}_{\rm z}$, we have that $\Omega=\omega/2$ (see also \cite{Chernin1993}, for a more complex and complete treatment of the interrelation of vorticity and angular momentum in galaxies). Then, the previous equations show a strict connection between vorticity, $\omega$, and angular velocity $\Omega$. 

It is then convenient to define the dimensionless $\alpha$-number as the ratio between the rotational and the gravitational term in Eq. (\ref{eqn:spher}):
\begin{equation}
\alpha(M)=\frac{L^2}{M^3RG}
\end{equation}
The above ratio, $\alpha$, is mass dependent. It has maximum values for galaxies\footnote{In the case of our galay it is $\simeq 0.1$.}, and decreases going toward clusters of galaxies.

In order to solve Eq. (\ref{eqn:wnldeq}), we should know how the term $\sigma^2-\omega^2$ depends on the density contrast. This can be done using the above outlined argument for rotation, namely the connection between angular momentum and shear, and recalling that Eq. (\ref{eqn:spher}) from which $\alpha$ was obtained is equivalent to Eq. (\ref{eqn:wnldeq0}) which is also equivalent to Eq. (\ref{eqn:nleq11}), and Eq. (\ref{eqn:wnldeq}). 

One may then calculate the same ratio between the gravitational and the extra term appearing in Eq. (\ref{eqn:nleq11}) or Eq. (\ref{eqn:wnldeq}) thereby obtaining
\begin{equation} \label{eq:nonlinf}
\frac{\sigma^2-\omega^2}{H^2_0}=-\frac{3}{2}\frac{\alpha \Omega_{\rm m,0}}{a^3}\delta.
\end{equation}

As shown in \citep{DelPopolo2013b} this is a reasonable assumption, and it was also used in \citep{DelPopolo2013a,DelPopolo2013b,Pace2014,Mehrabi2017}. 
%
%
}

In \citep{DelPopolo2013a,DelPopolo2013b} it was obtained calculating the threshold of collapse $\delta_c$, and matching it with $\delta_c$ obtained by \citep{Sheth1999}. 
%
}

Then the nonlinear equation to solve is obtained substituting Eq. (\ref{eq:nonlinf}) into Eq. (\ref{eqn:wnldeq})
\begin{equation}\label{eqn:wnldeqq}
  \delta^{\prime\prime}+\left(\frac{3}{a}+\frac{E^\prime}{E} \right)
\delta^\prime-\frac{4}{3}\frac{\delta^{\prime 2}}{1+\delta}-
  \frac{3}{2}\frac{\Omega_{\mathrm{m},0} (1-\alpha)}{a^5 E^2(a)}\delta(1+\delta)=0.
\end{equation}


The threshold of collapse, and the turnaround, can be obtained solving Eq. (\ref{eqn:wnldeqq}) following the method described in \citep{Pace2010}, or solving Eq. (\ref{eqn:wnldeq0}).

An important point is that Eq. (\ref{eqn:wnldeq0}), can be written in a more general form taking into account dynamical friction
 \citep{Kashlinsky1986,Kashlinsky1987,Lahav1991,Bartlett1993,
 AntonuccioDelogu1994,Peebles1993,DelPopolo1998,DelPopolo1998a,DelPopolo2006b,Delpopolo2009,DelPopolo2019}
\begin{equation}\label{eq:coll}
 \ddot{R} = -\frac{GM}{R^2} + \frac{L^2(R)}{M^{2}R^3} + \frac{\Lambda}{3}R -
 \eta\frac{{\rm d}R}{{\rm d}t}\,,
\end{equation}
being $\eta$ the dynamical friction coefficient. 
Eq. (\ref{eq:coll}) can be obtained via Liouville's theorem \citep{DelPopolo1999}, and the dynamical friction force per unit mass, $\eta\frac{{\rm d}R}{{\rm d}t}$, is given in \citep{Delpopolo2009}(Appendix D, Eq. D5), and \citep{DelPopolo2006b}, Eq. 5).

In terms of shear, vorticity, and DE, the equation, similarly to Eq. (\ref{eqn:wnldeq0}), can be written as
\begin{equation} \label{eqn:wnldeqqq}
\ddot{R} = -\frac{GM_{\rm m}}{R^2} - \frac{GM_{\rm de}}{R^2}(1+3w_{\rm de})-\frac{\sigma^2-\omega^2}{3}R -\eta \frac{dR}{dt}.
\end{equation}

A similar equation (excluding the dynamical friction term) was obtained by several authors 
\citep[e.g.,][]{Fosalba1998a,Engineer2000,DelPopolo2013b}) and generalized to smooth DE models in 
\cite{Pace2019}.

Eq. (\ref{eqn:wnldeqqq}), and Eq. (\ref{eqn:wnldeq0}) differs for the presence of the dynamical friction term. As shown in several papers (e.g., \citep{AntonuccioDelogu1994,Delpopolo2009,DelPopolo2006b,DelPopolo2019} dynamical friction has a similar effect to that of rotation, and cosmological constant: it delays the collapse of a structure (perturbation). 
As shown in Fig. 1 of \citep{DelPopolo2017}, and in Fig. 11 of \citep{Delpopolo2009}, the magnitude of the effect of rotation, dynamical friction, and cosmological constant are of the same order of magnitude with differences of a few percent. 
Willingly, by means of the relation $\delta=\frac{2GM_{\rm m}}{\Omega_{m,0} H^2_0}(a/R)^3-1$, Eq. (\ref{eqn:wnldeqqq}) can be written in terms of $\delta$, similarly to Eq. (\ref{eqn:wnldeq}). 

%
%

Summarizing, the typical quantities of the SCM depends from shear, vorticity, and dynamical friction. 
Contrarily to \citep{Pavlidou2014}, and several other authors (e.g., \citep{Lopes2018,Bhattacharya2017}), the SCM results depend from the baryon physics, since shear, rotation, and dynamical friction depends from the mass, and the way structures formed. 

In the following, we specialize on the effect shear, and rotation has on the SCM, and how the TAR is modified.
To this aim, in this paper, dynamical friction will be neglected.

\section{Results}\label{sect:Results}

\begin{figure*}[!ht]
 \centering
 \includegraphics[width=22cm,height=6cm,angle=0]{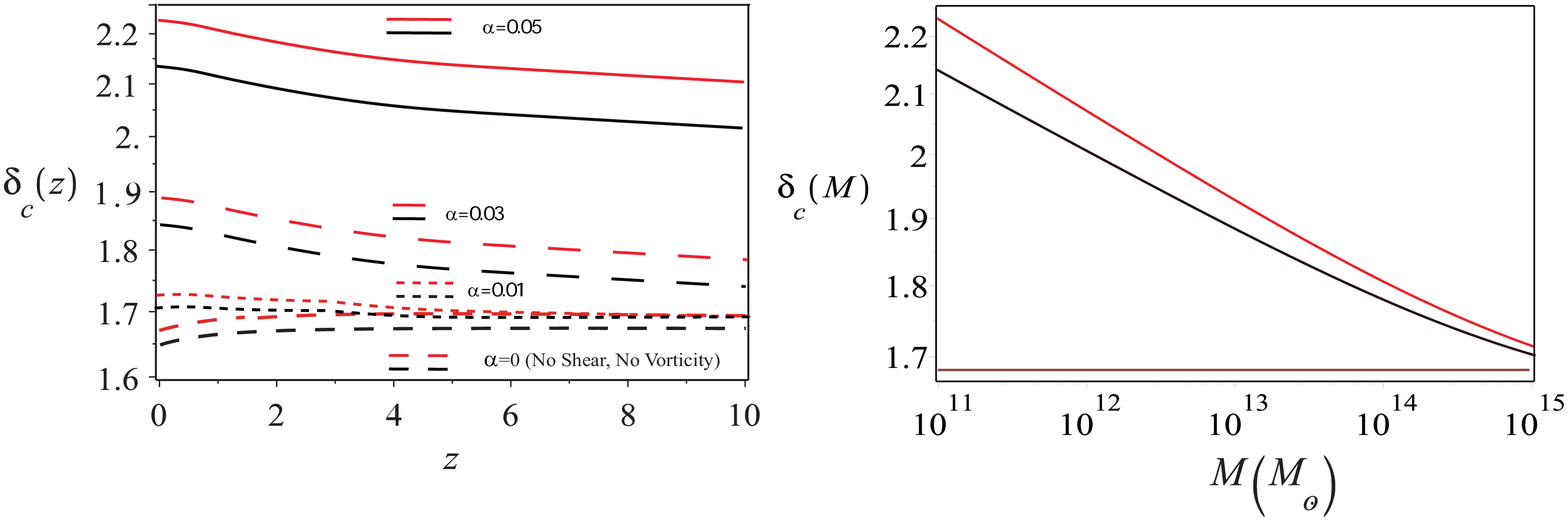} 
 \caption[justified]{The threshold of collapse $\delta_c$, as function of redshift, and mass.
In the left panel, we plot $\delta_c$ vs redshift. The red line represents the predictions of the ESCM for  $\delta_c(z)$ for the $\Lambda$CDM model. From top to bottom, the value of $\alpha$ varies from 0.05 (solid line) (galactic mass scale)
to 0.03 (long dashed line), 
to 0.01 (dotted line) (cluster mass scale), 
and to 0 (dashed line). Left panel: $\delta_c$ vs mass. The solid red line represents the result of the ESCM model, the red dashed line that of {the elliptictal collapse model of} \citep{Sheth2001}, and the brown line the $\Lambda$CDM expectation when no shear and rotation are taken into account.}
 \label{fig:comparison}
\end{figure*}

Several authors have studied the effect of shear \citep{Hoffman1986,Hoffman1989, Zaroubi1993}, and rotation \citep{DelPopolo1998,DelPopolo1999,DelPopolo2000,DelPopolo2001,DelPopolo2002,DelPopolo2013b,Pace2019} on the SCM and elliptical collapse  model. 
The effect of shear and rotation is that of slowing down the collapse \citep{Peebles1990,Audit1997,DelPopolo2001,DelPopolo2002}, changing turnaround epoch and collapse time, as well as the typical parameters of the SCM.
As a consequence also the mass function \citep{DelPopolo1999,DelPopolo2000,DelPopolo2013b,Pace2014,Mehrabi2017,DelPopolo2017,Pace2019}, the two-point correlation function, scaling relations like the mass-temperature \citep{DelPopolo2002a,DelPopolo2019}, and luminosity-temperature relation \citep{DelPopolo2005}, are modified.

\subsection{Threshold of collapse with shear, and rotation}

\begin{figure*}[!ht]
 \centering
\includegraphics[width=25cm,angle=0]{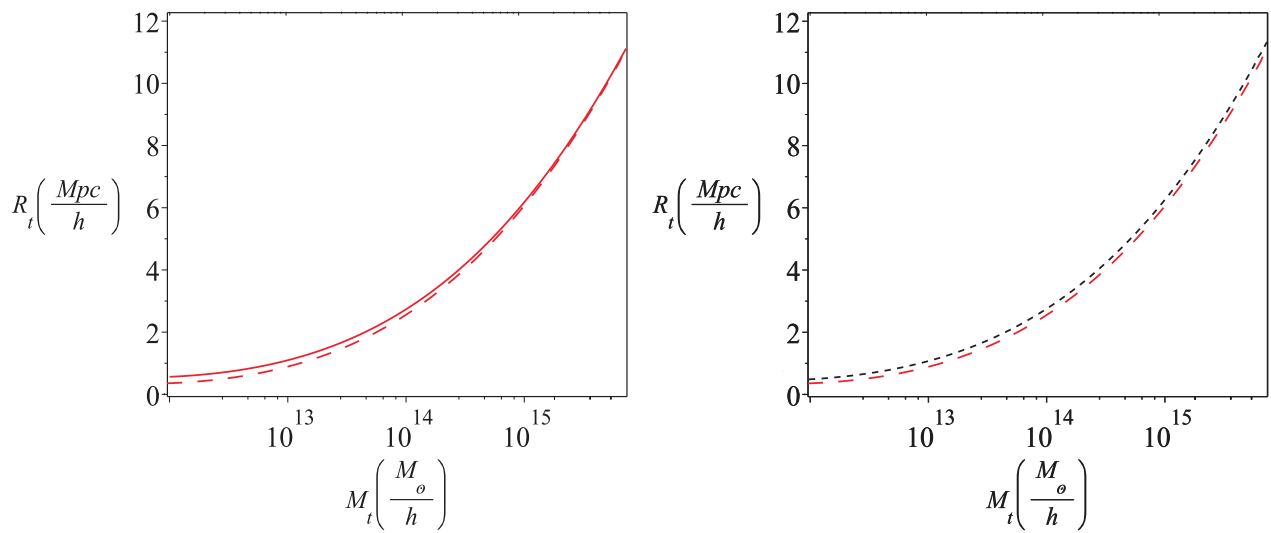}  
 \caption[justified]{Turnaround radius $R_{\rm t}$ vs mass, $M_{\rm t}$. Left panel: TAR predicted by the standard SCM (solid red line), and the ESCM (dashed red line)
for the $\Lambda$CDM model. Right panel: the result of the ESCM for the $\Lambda$CDM model (red dashed line), and that of the AS model (black dotted line). 
}
 \label{fig:comparison}
\end{figure*}

In order to show the effects of shear and rotation on the SCM predictions, we show, in Fig. 1, how the collapse threshold,  $\delta_c$, is modified. In the left panel, we show the dependence of $\delta_c$ from the redshift. The red line represents the predictions of the ESCM for  $\delta_c(z)$ for the $\Lambda$CDM model. From top to bottom, the value of $\alpha$ varies from 0.05 (solid line) corresponding to a mass $\simeq 10^{11} M_{\odot}$, to 0.03 (long dashed line), corresponding to a mass $\simeq 10^{13} M_{\odot}$, to 0.01 (dotted line), corresponding to a mass $\simeq 10^{15} M_{\odot}$, and to 0 (dashed line). 
We also show the predictions of the extended SCM in the case of one DE model, the Albrecht-Skordis \citep{Albrecht2000} (AS) DE model (black line). We chose to plot this model, because the other quintessence models considered in previous papers \citep{Pace2010,DelPopolo2013a} INV1 ($w_0=-0.4$), INV2 ($w_0=-0.79$), 2EXP ($w_0=-1$), CNR ($w_0=-1$), CPL ($w_0=-1$), SUGRA ($w_0=-0.82$) (see \citep{Pace2010,DelPopolo2013b})\footnote{$w_0$ is the value of $w$ nowadays} have a similar behavior to that of the AS model, and because the values of $\delta_c(z)$ predicted by them is located in the region between the $\Lambda$CDM and the AS model (see Fig. 4 of \citep{Pace2010}).  
The inclusion of the term proportional to $\sigma^2-\omega^2$ changes only the value of the linear overdensity parameter. The ratios with the $\Lambda$CDM model does not change. This means, that 
{if the $\delta_c$ relative to the $\Lambda$CDM is higher than $\delta_c$ relative to a given DE model, the same relationship will remain after taking into account the $\sigma^2-\omega^2$ term. This can be seen comparing the top left panel in Fig. 4 of \citep{Pace2010} with the top left panel in Fig. 2 of \citep{DelPopolo2013a}, and also looking at Fig. 1 of the present paper.}
%
%
The DE models similarly to the $\Lambda$CDM model are "retarded" (the collapse is slowed down) by the presence of the shear, and rotation term, 
and their values for $\delta_c(z)$ are all smaller than the $\delta_c(z)$ of the $\Lambda$CDM model obtained with the ESCM. This is expected, since at early times the DE amount is larger, and consequently the value of $\delta_c(z)$ must be lower to have objects collapsing.

\begin{figure*}[!ht]
 \centering
 \includegraphics[width=12cm,angle=0]{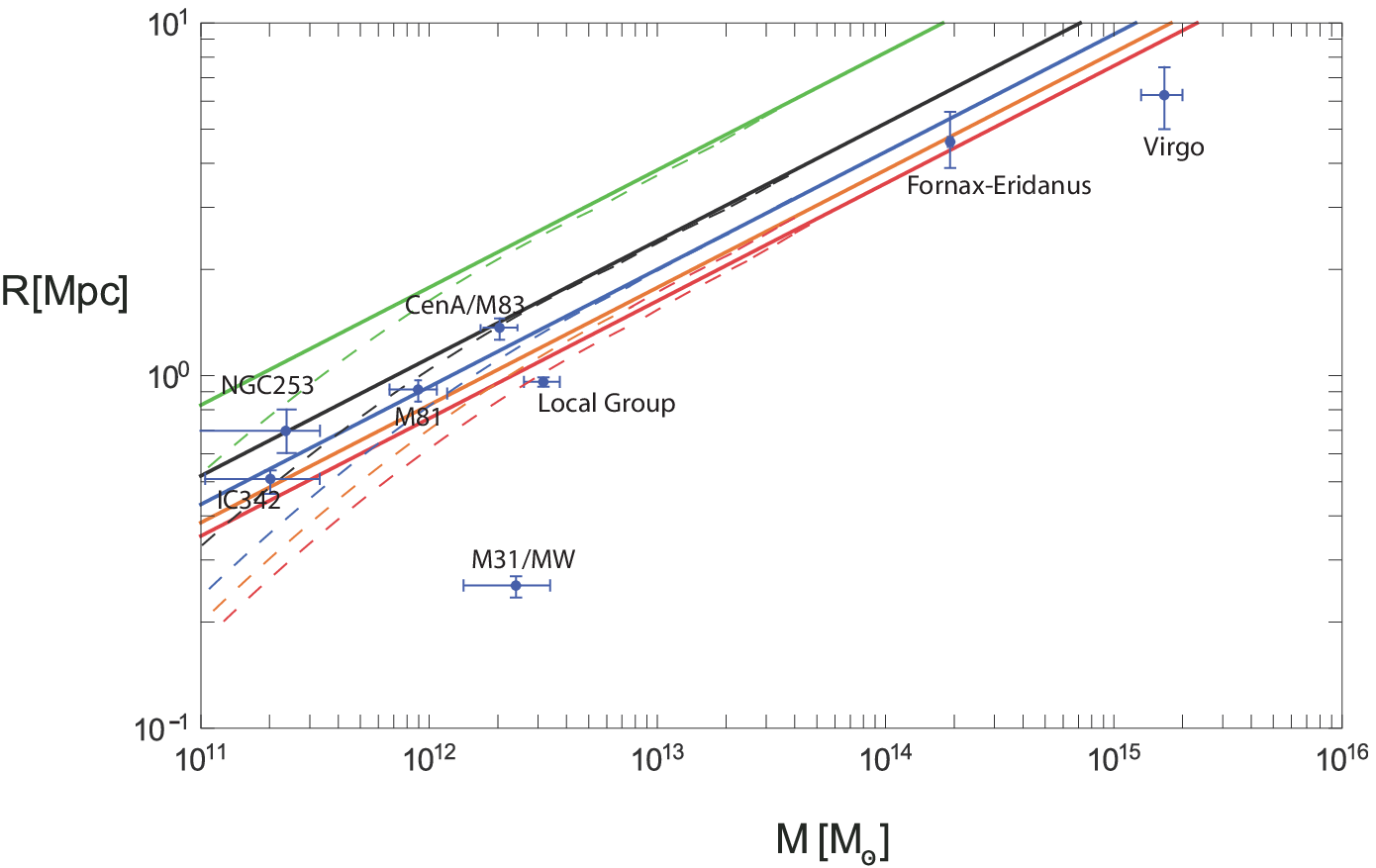}
 \caption[justified]{Mass-radius relation of stable structures for different $w$. The solid lines from top to bottom represent $w=-0.5$ (solid green line), -1 (black solid line), -1.5 (blue solid line) ,-2 (pink solid line), -2.5 (red solid line). The dashed lines are the same of the previous lines, but they are obtained using the ESCM model. The dots with error bars, are data as in \citep{Pavlidou2014a}.}
 \label{fig:comparison}
\end{figure*}

The plot also shows the different behavior of $\delta_c(z)$ in the case of no shear, and rotation ($\alpha=0$). In this case, $\delta_c$ has a weak dependence from redshift in the $z$ range $[0,2]$, and then becomes constant, with a value equal to that of the Einstein de Sitter model. Shear, and rotation ($\alpha \neq 0$), change the behavior of $\delta_c(z)$, as follows: 1. it assumes larger values with respect the case shear and rotation are not present; 2. the larger is $\alpha$, the larger is the difference between the values of $\delta_c(z)$; 3. $\delta_c(z)$ has a decaying behavior. The value of $\delta_c(z)$ for $\alpha=0.05$, is $\simeq 30\%$ larger than in the $\alpha=0$ case. In the right panel of Fig. 1, we plot $\delta_c(M)$ versus the mass. When shear and rotation are not present the value of $\delta_c$ is constant (brown line), while when they are taken into account the threshold $\delta_c$ becomes mass dependent, and is characterized by larger values at smaller masses, converging to the standard $\Lambda$CDM model constant value moving to the largest clusters. The dashed line shows the result of \citep{Sheth1999,Sheth2001}. 
The behavior of the threshold implies that less massive perturbations (e.g., galaxies) in order to form structures must cross a higher threshold than more massive ones. In terms of the peak formalism, and the peak height $\nu=\delta_c/\sigma(M)$\footnote{$\sigma(M)$ is the mass variance}, the angular momentum acquired by a peak is proportional to turnaround time, $t_{\rm ta}$, and anticorrelated with the peak height $j \propto t_{\rm ta} \propto \nu^{-3/2}$ \citep{Hoffman1986,Delpopolo2009,Polisensky2015}. 
Since low peaks acquire larger angular momentum than high peaks, they need a higher density contrast to collapse and form structures \citep{DelPopolo1998,DelPopolo2001,DelPopolo2002,Delpopolo2009,Ryden1988,Peebles1990,Audit1997}.
The cosmological constant has similar effects to that of angular momentum in collapse, slowing down it, but its effects vanishes at high redshift. 
%
\subsection{TAR, in the $\Lambda$CDM, and DE models}

\begin{figure*}[!ht]
 \centering
\includegraphics[width=18cm,angle=0]{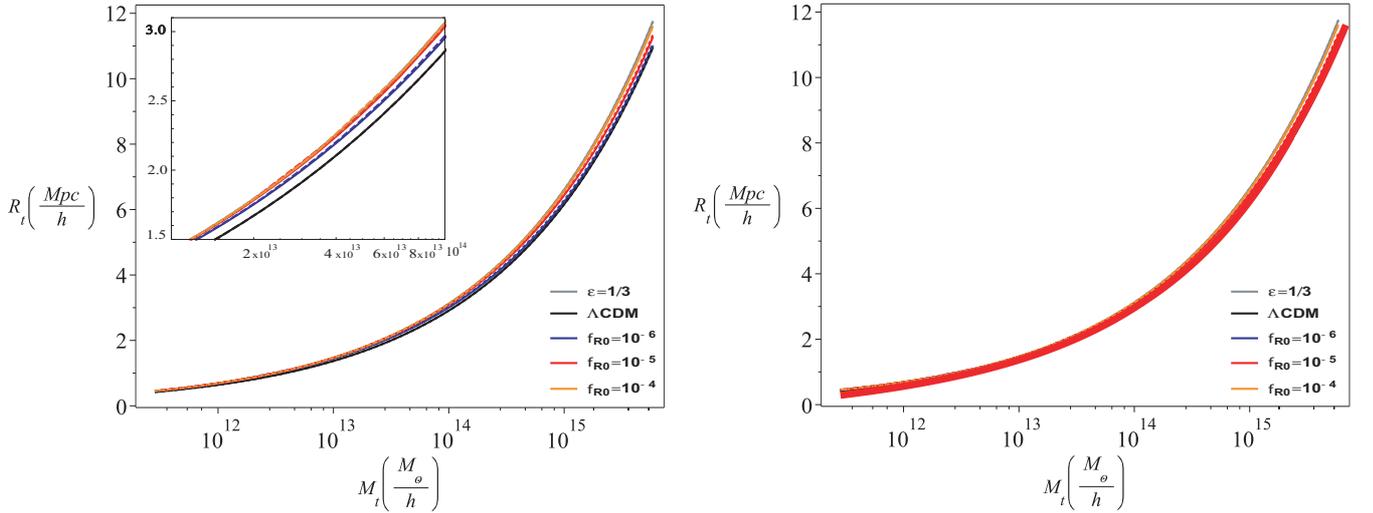} 
 \caption[justified]{Turnaround radius, $R_{\rm t}$ vs mass, $M_{\rm t}$. Left panel: the result obtained by \citep{Lopes2018} for $R_{\rm ta}$ vs mass for the $\Lambda$CDM model, the model with $\epsilon=1/3$, and $f(R)$ with $f_{\rm R0}=10^{-6}$, $10^{-5}$, and $10^{-4}$. Right panel: the same as in left panel compared with the prediction of the ESCM. The red band represents our prediction for the $\Lambda$CDM model, obtained with the ESCM, with the 68\% confidence level region.}
 \label{fig:comparison}
\end{figure*}

\begin{figure*}[!ht]
 \centering
\includegraphics[width=18cm,angle=0]{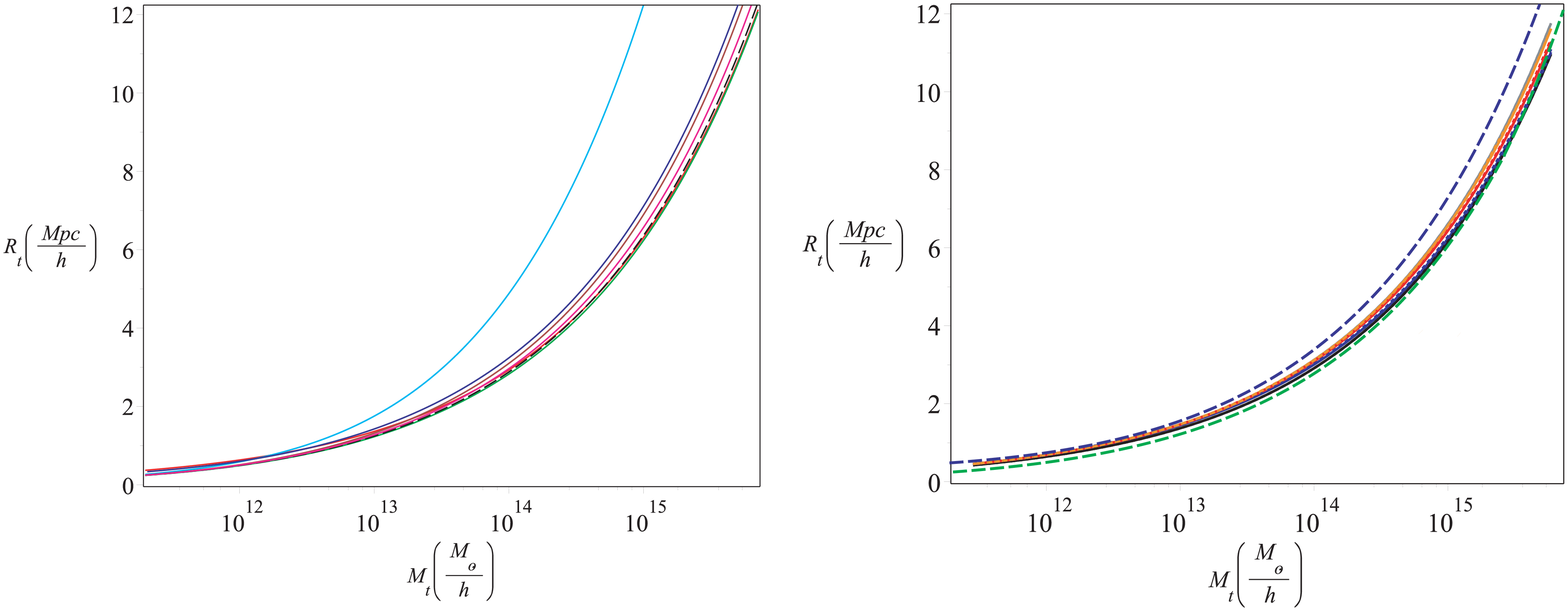}
 \caption[justified]{The turnaround radius, $R_{\rm t}$ vs mass, $M_{\rm t}$ obtained with the ESCM for the DE models. In the left panel, from top to bottom, the cyan, blue, brown, magenta, black, red, and green lines represent the INV1, INV2, SUGRA, $w_{09}$\footnote{Namely the model having $w=-0.9$}, AS, $\Lambda$CDM without shear, and rotation, and $\Lambda$CDM with shear, and rotation, respectively. In the right panel, the solid lines are the $\Lambda$CDM, and $f(R)$ models of \citep{Lopes2018} (similarly to the left panel of our previous figure), while the blue dashed line, and the green dashed lines, are the INV2 model, and the $\Lambda$CDM with shear, and rotation plotted in the left panel.}
 \label{fig:comparison}
\end{figure*}

Our main goal is to understand how shear and rotation changes the TAR. We start finding an expression for the TAR. {Just in order to compare with \citep{Pavlidou2014} and \citep{Pavlidou2014a} results, we calculate the maximum TAR, MTAR,
namely the radius of the surface where $\ddot R =0$}. Using Eq. (\ref{eqn:wnldeq0}), we obtain
\begin{equation} \label{eq:turnn}
R_{\rm ta}= \left [\frac{-3M}{4 \pi \rho_{\rm de}(1+3w)+(\sigma^2-\omega^2)} \right]^{1/3}
\end{equation}
In the case of the $\Lambda$CDM model, ($w=-1$), no shear, and rotation, Eq. (\ref{eq:turnn}) reduces to
\begin{equation} \label{eq:turn1}
R_{\rm ta}= \left [\frac{3GM}{\Lambda} \right]^{1/3}
\end{equation}
which is the same as that obtained by \citep{Pavlidou2014}. In the case of DE models, with no shear, and rotation, Eq. (\ref{eq:turnn}) reduces to 
\begin{equation} \label{eq:pavl}
R_{\rm ta}= \left [\frac{-3M}{4 \pi \rho_{\rm de}(1+3w)} \right]^{1/3}
\end{equation}
which is the same as that obtained by \citep{Pavlidou2014a}. 
In other terms, the turn around obtained by \citep{Pavlidou2014a} is a peculiar case of the more general form given by Eq. (\ref{eq:turnn}). 
%
%

{In the paper, we will get and plot the TAR, not MTAR, since we compare with \citep{Lopes2018}, which calculated the TAR.}

When taking into account shear, and rotation, the TAR is smaller than expected from \citep{Pavlidou2014a} result, as shown in Fig. 2. In the left panel of Fig. 2, we plot the TAR predicted by the standard SCM (solid red line), and the ESCM (dashed red line)
for the $\Lambda$CDM model. As expected, similarly to the threshold of collapse, $\delta_c$, when shear and rotation are taken into account, the collapse is slowed down and the turn around is smaller. 
The differences between the standard SCM predictions, and the ESCM, are larger at smaller masses, because smaller mass objects have larger rotation.
In the right panel of Fig. 2, we compare the result of the ESCM for the $\Lambda$CDM model (red dashed line) with that of the AS model (black dotted line). 
The AS model has a slightly larger TAR with respect to the $\Lambda$CDM model. 
\citep{Pavlidou2014a} tried to constrain the equation of state by means of Eq. (\ref{eq:pavl}), comparing the predicted TAR at different $w$ with mass and radii of small $z$ cosmic structures. 
In this way is possible to constrain the equation of state parameter $w$ at $z \simeq 0$ ($w_0$)\footnote{Many of the DE models can be described by an equation of state parameter, $w(a)$, depending on its value nowadays, $w_0$, that at matter-radiation equality epoch, and some other parameters at the same epoch (see Eq. 23 of \citep{Pace2010}). In order to constrain the evolution of the equation of state parameter, $w$, it is necessary to use structures at high redshift. }. They used data for: Milky Way (MW), M81/M82 group, Local Group, Virgo cluster, Fornax-Eridanus group. \citep{Peirani2008} took into account the cosmological constant and obtained different values for M31/MW, M81, Virgo, and also obtained the R-M values for CenA/M83, IC342, NGC253. We used the data obtained by \citep{Peirani2008}, and that of \citep{Pavlidou2014a} for Local Group, and Fornax-Eridanus. 

In Fig. 2, the solid lines are obtained from the equation of TAR (Eq. \ref{eq:pavl}) not taking into account shear, and rotation. The dashed lines are the corrections obtained when shear, and rotation are taken into account. The lines correspond to $w=-2.5$, -2, -1.5, -1, -0.5, from bottom to top. The region of the parameter space above each line gives the range of $w$ for which no stable structures should exist. \citep{Pavlidou2014a} discussed some constraints to $w$, based on the highest mass objects. Here we want just to stress that at masses smaller than $10^{14} M_{\odot}$ the TAR is modified by the presence of shear, and rotation. As a consequence, structures at smaller masses can give different constraints to $w$. 

For example, the Local Group gives the constrain $w \geq -2.5$, while if shear, and rotation are not taken into account, $w$ is more negative.
{The other constraints are reported in Table 1}.
%
%
%
{Therefore, including shear and rotation can help to constrain the equation of state of DE by means of the data of galaxies. Furthermore, 
can only marginally satisfy the data of NGC253 and M81, the standard $\Lambda$CDM model can also be strongly constrained by galactic data.}

\subsection{Comparison with TAR in $f(R)$ theories}

\citep{Lopes2018} investigated the evolution of the TAR, and its dependence from structure mass in General Relativity (GR), and in the $f(R)$ theories. Their Eq. (3.3) takes into account the effects of MG through the parameter $\epsilon(a,k)$, where $k$ is the (angular) wavenumber. When $\epsilon(a,k)$ is zero GR is recovered, and their Eq. (3.3) is equal to our Eq. (\ref{eqn:wnldeq}) without shear, and rotation. In other terms, \citep{Lopes2018} includes the effects of the $f(R)$ theories but not that of shear, and rotation, which changes the behavior of the main quantities of the SCM. As a consequence,
their Figs. 1, 3 predict a monotonic increase for $\delta_c(z)$ as 
in the $\Lambda$CDM, and DE models without shear, and rotation, as shown in \citep{Pace2010}, and in the left panel of our Fig. 1 (case $\alpha=0$). Shear, and rotation modify this behavior. 
Similarly, their Fig. 4 shows that $\delta_c(M)$, differs from the $\Lambda$CDM predictions by $\simeq 1\%$.  
As shown by \citep{Sheth2001}, and by \citep{DelPopolo2013a,DelPopolo2013b,Pace2014,Mehrabi2017,DelPopolo2017}, 
in order to have a mass function reproducing simulations, the threshold 
must be: 1. mass dependent, and 2. a decaying function of mass.  
For precision's sake, in the excursion set formalism, the halo statistics, and the mass function, depends from the statistical properties of  the average overdensity, $\overline{\delta}_{\rm R_f}$, within a window of a given radius, $R_f$. $\overline{\delta}_{\rm R_f}$ vs $R_f$ is a random walk \citep{Bond1991,DelPopolo2017}, and haloes form when the random walk crosses the threshold, $\delta_c$, dubbed barrier \citep{Sheth2001,DelPopolo2007a,DelPopolo2017}. A mass function that reproduces simulations and observations must have a mass dependent, and monotonically decreasing $\delta_c$, as in \citep{Sheth2001,Sheth2002}, 
and \citep{DelPopolo1998,DelPopolo1999,DelPopolo2017}. This shows that the results of \citep{Lopes2018} cannot reproduce the mass function obtained in simulations, and observations, because the effects of 
shear, and rotation
are not taken into account. 
~\\

Another important point is the use of the TAR to disentangle between GR and MG, discussed by several authors (e.g., \citep{Lopes2018}).
In the following, we will compare \citep{Lopes2018} results with the prediction of the ESCM for the TAR in the $\Lambda$CDM, and DE models. To this aim, we compared in Fig. 4, their result concerning the $R_{\rm t}$ vs mass with the $\Lambda$CDM model predictions in the ESCM. In the figure, we plot the result obtained by \citep{Lopes2018} for $R_{\rm t}$ vs mass for the $\Lambda$CDM model, the model with $\epsilon=1/3$, and $f(R)$ with $f_{\rm R0}=10^{-6}$, $10^{-5}$, and $10^{-4}$. The red band represents our prediction for the $\Lambda$CDM model, obtained with the ESCM, with the 68\% confidence level region. 
{In order to calculate the 68\% confidence level, we used a Monte Carlo simulation.  
In short, this approach involves the repeated simulation of samples within the probability density functions of the input data. In the first step, one looks for the sources of uncertainty, and identifies the probability density functions (PDFs) that gives a good fit to those the data sources.
For all data inputs identified as sources of uncertainty, Monte Carlo simulations are run, and the resulting simulations are applied to models (equations), in order to get the distribution of the quantity of interest. Finally one can get the confidence intervals needed.   
In our case, in order to obtain the TAR, one has to solve Eq. (\ref{eqn:wnldeq0}), or Eq. (\ref{eqn:wnldeqq}), using the condition $\dot R=0$ at turnaround or equivalently a relation between 
$\delta$, $\delta^{'}_t$, and at turn around (see eq. 3.8 of \citep{Lopes2018}). If for example, we consider Eq. (\ref{eqn:wnldeq0}), we may write it as
\begin{equation}
\dot R^2=0=2 \frac{GM}{R}-\frac{4 \pi \rho_{\rm de}}{3} +\frac{2 H^2_0 \Omega_{m,0}}{2 a^3 G} 
\int_{R_i}^{R_t} \frac{L^2}
{M^3} \delta dR
\end{equation}
where $R_i$ is the initial radius of the perturbation, related to $z_i$ given in the appendix.
One can get numericaly the relation between $R_t$, and $M_t$. In order to obtain the confidence interval by means of the Monte Carlo method, we observe that $R_t$ depends from mass, overdensity, and angular momentum.
Angular momentum has a lognormal distribution (\citep{Vitviska2002,Romanowsky2012}), and the overdensity has a Gaussian distribution. By means of Monte Carlo simulations we can get the 68\% confidence region. Another way, is to get $R_t$ in terms of the overdensity, since the angular momentum is proportional to $M^{5/3}$ (\citep{Ryden1987,Ryden1988,Catelan1996a,Romanowsky2012}, and $M=\frac{4 \pi}{3} R^3 \overline{\rho} (1+\delta)$. Recalling that $\delta$ has a Gaussian distribution, we may get the confidence region by means of the monte carlo method.
}
The plot shows that the $\Lambda$CDM model prediction is similar to that of the MG theories. Similarly to \citep{DelPopolo2019}, in which we showed that the mass-temperature relation {is not a good probe to test gravity theories beyond GR}, and to disentangle between GR, and $f(R)$ theories, here we have a similar result: the TAR is not a good probe to disentangle between GR, and $f(R)$ theories. In Fig. 5, we compare the predictions of some of the quintessence DE models previously cited 
to the same $f(R)$ models of \citep{Lopes2018} plotted in Fig. 4. In the left panel of Fig. 5, we plot the predictions of the DE models for the TAR, obtained with the ESCM. From top to bottom, the cyan, blue, brown, magenta, black, red, and green lines represent the INV1, INV2, SUGRA, $w_{09}$\footnote{Namely the model having $w=-0.9$}, AS, $\Lambda$CDM without shear, and rotation, and $\Lambda$CDM with shear, and rotation, respectively.

In the right panel, we compare the \citep{Lopes2018} prediction with that of the DE models, plotted in the left panel. In order to have a more readable figure, we plotted just the INV2 (blue dashed line), and the $\Lambda$CDM with shear, and rotation (green dashed line) plotted in the left panel. These two lines contain all the other DE models (except INV1). The solid lines are the $\Lambda$CDM, and $f(R)$ models of \citep{Lopes2018}. 
Similarly to what happens in Fig. 4, the DE models have similar predictions for the TAR to that of \citep{Lopes2018}, showing that the quoted radius cannot be used to disentangle DE, and $f(R)$ theories predictions. 

%
%
\begin{table}
\caption{The ruled out ranges of $w$ based on the ESCM model.}
 \label{table1}
 \begin{tabular}{@{}lccc}
  \hline
  Stable structure & $M$ ($M_{\odot}$) & $R$ (Mpc) & range of $w$ ruled out \\
  \hline
  M31/MW & $(2.4 \pm 1.0) \times 10^{12}$ & $0.25 \pm 0.02$ & $ w >  -2.5$ \\
  M81 & $(8.9 \pm 2.2) \times 10^{11}$ & $0.90^{+0.06}_{-0.05}$ & $w \ge -1$ \\
  IC342 & $2.0^{+1.2}_{-1.0} \times 10^{11}$ & $0.51^{+0.03}_{-0.05}$ & $w \ge -1$ \\
  NGC253 & $2.34^{+0.99}_{-1.34} \times 10^{11}$ & $0.70 \pm 0.10$ & $w \ge -0.5$ \\
  CenA/M83 & $2.00^{+0.42}_{-0.33} \times 10^{12}$ & $1.37^{+0.08}_{-0.11}$ & $w \ge -1$ \\
  Local Group & $3.14^{+0.57}_{-0.54} \times 10^{12}$ & $0.96 \pm 0.03$ & $w \ge -2.5$ \\
  Fornax-Eridanus & $1.92 \times 10^{14}$ & $4.60^{+1.0}_{-0.72}$ & $w \ge -1.5$ \\
  Virgo & $1.67^{+0.31}_{-0.35} \times 10^{15}$ & $6.5^{+1.0}_{-1.5}$ & $w \ge -2.5$ \\
  \hline
 \end{tabular}
\end{table}

\section{Conclusions}\label{sect:conclusions}

In the present work, we discussed how shear, and rotation change the TAR, and 
some of the parameters of the SCM. We used an extended SCM taking into account the effects of shear and vorticity to determine the $R_{\rm t}-M_{\rm t}$, in $\Lambda$CDM, and in DE scenarios. From the condition that the radial acceleration is zero, we got a formula for the maximum TAR more general than those found in literature. Shear, and rotation reduce the value of the TAR, particularly at the galactic scales where rotation is larger. As a consequence, using the $R_{\rm t}-M_{\rm t}$ relationship, and data from stable structures to constrain the $w$ parameter, one gets smaller values (absolute value) of the quoted parameter for structures having masses smaller than $10^{13} M_{\odot}$. 
We also compared the $R_{\rm t}-M_{\rm t}$ relationship obtained for $\Lambda$CDM, and DE scenarios
with the prediction of the $f(R)$ theories, calculated by \citep{Lopes2018}. The result of the comparison shows that the $R_{\rm t}-M_{\rm t}$ relationship in the $f(R)$ models are practically identical to that of 
the $\Lambda$CDM model, and DE scenarios. This implies that the $R_{\rm t}-M_{\rm t}$ relationship is not a good probe to disentangle between GR, and DE models predictions.


\appendix

\section{f(R) theories and the Hu-Sawicki model }\label{sect:MG}

In this section, we summarize the MG theory in which framework \citep{Lopes2018} determined the TAR, which we compared to our model. 

\citep{Lopes2018} used the Hu-Sawicki model, which is one peculiar model in the $f(R)$ theories, 
whose action may be written as follows:
\begin{equation}
S =  \int{d^4 x \sqrt{-g} \left[ \frac{R+f(R)}{16\pi G} + \mathcal{L}_{m} \right]} \, ,\label{eqn:action}
\end{equation}
being $\mathcal{L}_{m}$ is the matter Lagrangian density. The $\Lambda$CDM model is recovered, in the case $f(R)= -2 \Lambda$.

Einstein modified equations are:
\begin{equation}
G_{\alpha \beta} + f_R R_{\alpha \beta} - \left(\frac{f}{2} - \Box f_R \right)g_{\alpha \beta}- \nabla_{\alpha} \nabla_{\beta} f_R = 8\pi G \, T_{\alpha \beta} \, , \label{egm}
\end{equation}
where $G_{\alpha \beta}$ is the Einstein tensor, $f_{R} \equiv \frac{d f(R)}{dR}$, and $T_{\alpha \beta}$ is the energy-momentum tensor. 

The $f(R)$ theories are 
characterized by the presence of the so called Chameleon screening mechanism, having a scalar field mass depending on the local density. In low density fields, where deviations from GR are maximized, the scalar degree of freedom is long ranged, and in high density ones happens the opposite. 

\citep{Lopes2018} used the Hu-Sawicki \citep{Hu2007} model, characterized by the following functional form 
\begin{equation}
 f(R) = -m^{2}\frac{c_1(R/m^2)^n}{1 + c_2(R/m^2)^n}\,,
\end{equation}
where $m$, $c_1$, $c_2$, and $n$, ($n>0$) are free parameters, with  $m^2 = H_0^2\Omega_{\rm m,0}$\footnote{It has dimensions of mass squared}. $c_1$ and $c_2$ can be determined by requiring that 
\begin{equation}
 \frac{c_1}{c_2} \approx 6\frac{\Omega_{\Lambda,0}}{\Omega_{\rm m,0}}\,.
\end{equation}
in the large curvature regime ($R/m^2\gg 1$), $f(R)\approx -2\Lambda$. 

$f_R$ today indicates the strength of gravity modifications:
\begin{equation}
 f_{R0} = -n\frac{c_1}{c_2^2}\left[\frac{\Omega_{\Lambda,0}}{3(\Omega_{\rm m,0} + 4\Omega_{\Lambda,0})}\right]^{n+1}\,.
\end{equation}
The range of the scalar degree of freedom is given by the comoving Compton wavelength $\lambda_c \equiv \frac{1}{{m_{f_R}}}$, where
\begin{equation}
m_{f_R}\approx  \frac{1}{\sqrt{3 f_{RR}}} \, .
\label{eq:massadocampescalaprox}
\end{equation}

The $f(R)$ theories are characterized by a scalar field coupling  
to all forms of matter having an energy-momentum tensors with non-zero traces.
Consequently, apart the GR gravitational force, the conformal coupling between matter and field produces a fifth force proportional to the field gradient (see \citep{DelPopolo2019}). 

The modified Poisson equation in comoving coordinates is given by:
\begin{equation}
\nabla^2 \Phi = \frac{16 \pi G}{3}a^{2} \delta \rho_{\rm m} - \frac{a^{2}}{6} \delta R(f_R) \, . \label{potorig}
\end{equation}
where $\delta \rho_m = \rho_m-\bar{\rho}_m$, and
$\delta R = R-\bar{R}$\footnote{A bar indicates a spatial average}.

In Fourier space, Eq. (\ref{potorig}) can be written as:
\begin{equation}
- k^2 \Phi = \left[1+ \epsilon(a,k) \right] 4 \pi G a^2 \delta \rho_m \, , 
\label{eq:eqpoissonfrconjM}
\end{equation}
with:
\begin{equation}
\epsilon(a,k) \equiv \frac{1}{3} \left(\frac{k^2}{a^2 m_{f_R}^2 + k^2}\right) \, .
\label{eq:defepsilonM}
\end{equation}
The modifications to GR are contained in the term $\epsilon(a,k)$. When this term vanishes we are left with GR. 

In order to have a closed system Poisson equation must be coupled with {the relativistic fluid equations for the matter density fields}.


{
The equation to be solved to get $\delta(r)$ is Eq. 3.3 of \citep{Lopes2018}
\begin{eqnarray}
\delta''+\left({3\over a}+{E'\over E}\right)\delta'-{4{\delta'^{2}}\over {3 (1+\delta)}}&=& \nonumber\\ 
\frac{3(1+\delta)}{2 \, E^2 \,2 \pi^2}  \,\Omega_{m0}\,a^{-5}\int_0^{\infty} dk \, k^2 [1+\epsilon(k,a)] \delta(k,a) \frac{\sin(kr)}{k r} \,.
\label{deltanonlinearx}
\end{eqnarray}
In the case $\epsilon(k,a)=0$, this reduces to our Eq. (\ref{eqn:wnldeq}) with $\sigma=\omega=0$.
In order to obtain $\delta(r)$, it is necessary to have an initial density profile, $\delta_i(r)$ to evolve from an early time, that we choose as, \citep{Lopes2018}, $z_i=500$, because at that time GR and MG are unistinguishable. In this paper, we used a top-hat-like profile, the Tanh profile of \citep{Lopes2018}
\begin{equation}
\delta_i(r)=\frac{\delta_i,0}{2}[1-\tanh{\frac{r/r_b-1}{s}}]
\end{equation}
where $s$ is the steepnes of the transition, and $r_b$ is the size of the top-hat-like function.
In the present paper, in order to obtain $\delta(r)$, we will solve our Eq. (\ref{eqn:wnldeq}) with the previous top-hat-like profile.
}

%

%
%

%

\end{document}